\documentclass{appolb}
\usepackage{graphicx}

\usepackage{hyperref}
\usepackage{amsmath}
\usepackage{amssymb}
\usepackage{color}

\def\be{\begin{eqnarray}}
\def\ee{\end{eqnarray}}
	
\allowdisplaybreaks

\begin{document}
\title{Chiral-scale effective field theory for dense and thermal systems%
\thanks{Presented at Excited QCD 2026 Workshop, Granada, Spain, 2026.01.08-01.14.}%
}
\author{Yong-Liang Ma
\address{School of Frontier Sciences, Nanjing University, Suzhou 215163, China}
%
}
\maketitle
\begin{abstract}
In this contribution, I will present some properties of nuclear matter (NM) by using the chiral-scale effective field theory that is anchored on the chiral, scale and hidden local flavor symmetries of QCD. We show that the sound velocity (SV) of the compact star matter can saturate the conformal limit, the SV exhibits a peak configuration in the intermediate density. To extend the chiral-scale effective field theory to both dense and tnermal systems, we setup a chiral-scale density counting (CSDC) rule and explore the contributions up to $\mathcal{O}(k_c^{12})$.  
\end{abstract}
  
\section{Introduction}

Chiral effective field theories (EFTs) have been widely and successfully used in nuclear physics~\cite{Machleidt:2011zz,Holt:2014hma,Ma:2019ery}. The advantage of these approaches is that they are anchored on the chiral symmetry so that are closer to QCD than phenomenological models. Normally, in chiral EFTs, the pions and lowest-lying vector mesons other than nucleons can be consistently included. However, since the scalar mesons are integrated out in the nonlinear realization of chiral symmetry, the iso-scalar scalar meson $\sigma$, the significant degree of freedom responsible for the attractive force between nucleons is missing. This flaw weakens the feasibility of chiral EFTs in nuclear physics, especially in the mean field approach.

The trace anomaly of QCD provides a source of the $\sigma$ meson in hadronic models~\cite{Schechter:1980ak}. In 2012, Crewther and Tunstall suggested an approach to include the $\sigma$ meson in the chiral EFT~\cite{Crewther:2013vea} by conjecturing that QCD has an infrared fixed point (IRFP) and the theory is slight away from it. In this approach, the $\sigma$ meson is the Nambu-Goldstone boson (NGB) of the scale symmetry breaking---the dilaton---with small mass arising from the deviation from the IRFP that breaks scale symmetry explicitly. The existence of the IRFP is supported by some literature, e.g., Refs.~\cite{Cui:2019dwv,Deur:2025rjo}.

In this contribution, I will present some properties of NM by using the chiral-scale effective field theory systematized in Ref.~\cite{Li:2016uzn} that,
%
in addition to nucleons, includes the lightest scalar meson as NGB of scale symmetry breaking, pions as the NGBs of chiral symmetry breaking and the lowest-lying vector mesons as the gauge boson of hidden local flavor symmetry.

\section{Chiral-scale effective field theory}

For the present purpose, we decompose the Lagrangian in terms of the pure meson part \(\mathcal{L}_M\) and baryon part \(\mathcal{L}_B\) as
\begin{equation}
	\mathcal{L}=\mathcal{L}_M+\mathcal{L}_B ,
	\label{eq:LbsHLS}
\end{equation}
where 
\be
\mathcal{L}_M & = & \ f^2_\pi\Phi^2{\rm Tr}\left(\hat{\alpha}^\mu_\perp\hat{\alpha}_{\mu\perp}\right) +\frac{m^2_\rho}{g^2_\rho}\Phi^2{\rm Tr}\left(\hat{\alpha}^\mu_\parallel\hat{\alpha}_{\mu\parallel}\right)  -\frac{1}{2g^2_\rho}{\rm Tr}\left(V_{\mu\nu}V^{\mu\nu}\right) \nonumber \\
& &{} +\frac{1}{2}\left(\frac{m^2_\omega}{g^2_\omega}-\frac{m^2_\rho}{g^2_\rho}\right)\Phi^2{\rm Tr}\left(\hat{\alpha}^\mu_\parallel\right){\rm Tr}\left(\hat{\alpha}_{\mu\parallel}\right) -\frac{1}{2g^2_0}{\rm Tr}\left(V_{\mu\nu}\right){\rm Tr}\left(V^{\mu\nu}\right) \nonumber\\
& &{} +\frac{1}{2}\partial_\mu\chi\partial^\mu\chi+\frac{f^2_\pi}{4}\Phi^2{\rm Tr}\left(\mathcal{M}U^\dagger+U\mathcal{M}^\dagger\right)  +h_5\Phi^4+h_6\Phi^{4+\beta'}, \label{eq:LM}\\
\mathcal{L}_B & = & \bar{N} i \gamma_\mu D^\mu N-m_N \Phi \bar{N} N \nonumber\\
& &{} +\left[g_A C_A+g_A\left(1-C_A \right)\Phi^{\beta^{\prime}}\right] \bar{N} \hat{\alpha}_{\perp}^\mu \gamma_\mu \gamma_5 N \nonumber\\
& &{} +\left[g_{V_\rho} C_{V_\rho}+g_{V_\rho}\left(1-C_{V_\rho}\right) \Phi^{\beta^{\prime}}\right] \bar{N} \hat{\alpha}_{\|}^\mu \gamma_\mu N \nonumber\\
& &{} +\frac{1}{2}\left[g_{V_0} C_{V_0}+g_{V_0}\left(1-C_{V_0}\right) \Phi^{\beta^{\prime}}\right] {\rm Tr}\left[\hat{\alpha}_{\|}^\mu\right] \bar{N} \gamma_\mu N . \label{eq:LB}
\ee
For the notations and conventions, we refer to Refs.~\cite{Zhang:2024sju,Zhang:2024iye}

\section{The pseudoconformal structure}

The chiral-scale EFT is a framework where chiral symmetry is realized in a nonlinearly and the trace anomaly effect is involved through dilaton compensator. Therefore, it is a natural approach for considering the medium effect on the parameters in the sense of Brown-Rho scaling~\cite{Brown:1991kk}. For the explicit density dependence of the medium modified parameters, since we do not have any prior information from underlying QCD, we take the lessons from skyrmion crystal approach which predicts that the medium modified parameters, such as $f_\pi^\ast, m_N^\ast, m_\rho^\ast$, and so on, first decrease with density and then stay as constants after certain density characterized by the transition from skyrmion to half-skyrmion phase~\cite{Ma:2013ooa}. 

\begin{figure}[htb]
\centerline{%
\includegraphics[width=6.0cm]{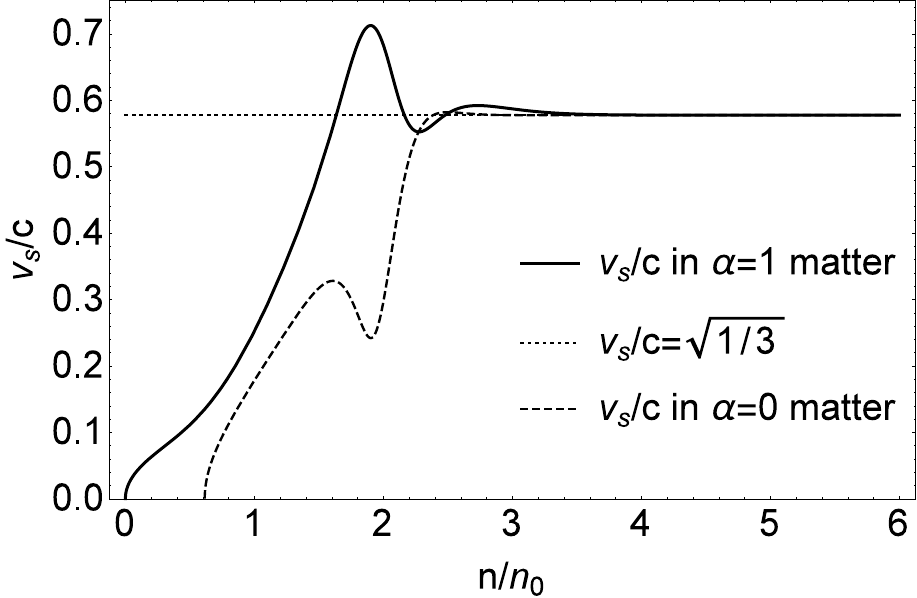}\quad \includegraphics[width=6.2cm]{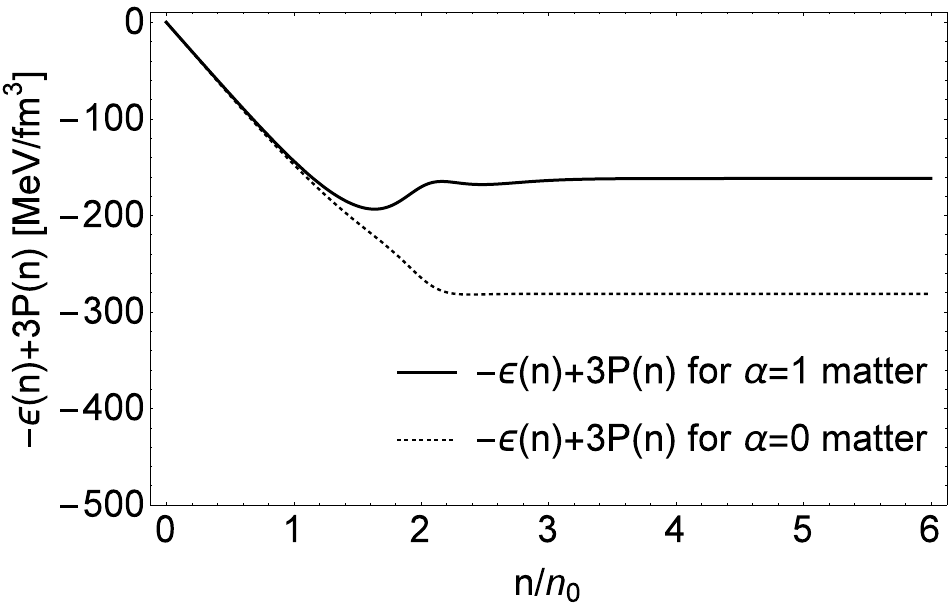}
}
\caption{Sound velocity (left panel) and TEMT (right panel) vs density for symmetric matter ($\alpha=0$) and neutron matter ($\alpha=1$). }
\label{fig:PC}
\end{figure}

By tuning the scaling parameter with respect to the NM properties around saturation density $n_0$, we found that the SV can saturate the conformal limit after the topology change located at density $n_{1/2}$~\cite{Paeng:2017qvp}, as shown in Fig.~\ref{fig:PC}. Note that, although the SV saturates the conformal limit, it does not mean that the theory is scale invariant, since its trace of the energy momentum tensor (TEMT) is a nonzero density independent constant (as shown Fig.~\ref{fig:PC}) that does not contribute to SV. 

The data from nuclear physics and heavy-ion collisions constrain $2.0n_0 \lesssim n_{1/2} \lesssim 4n_0$~\cite{Ma:2018xjw}. Therefore, the conformal SV can appear in the cores of massive NSs. This discovery is in stark contrast to the belief before that the conformal SV cannot appear at density below the perturbative QCD applicable, since otherwise the massive NS around two times solar mass cannot be obtained~\cite{Tews:2018kmu}. Although the discovery is novel when it was just found, it is observed in various models recently (see. e.g, Ref.~\cite{Kapusta:2021ney}).

\section{Peak of sound velocity}

\begin{figure}[htb]
	\centerline{%
		\includegraphics[width=6.0cm]{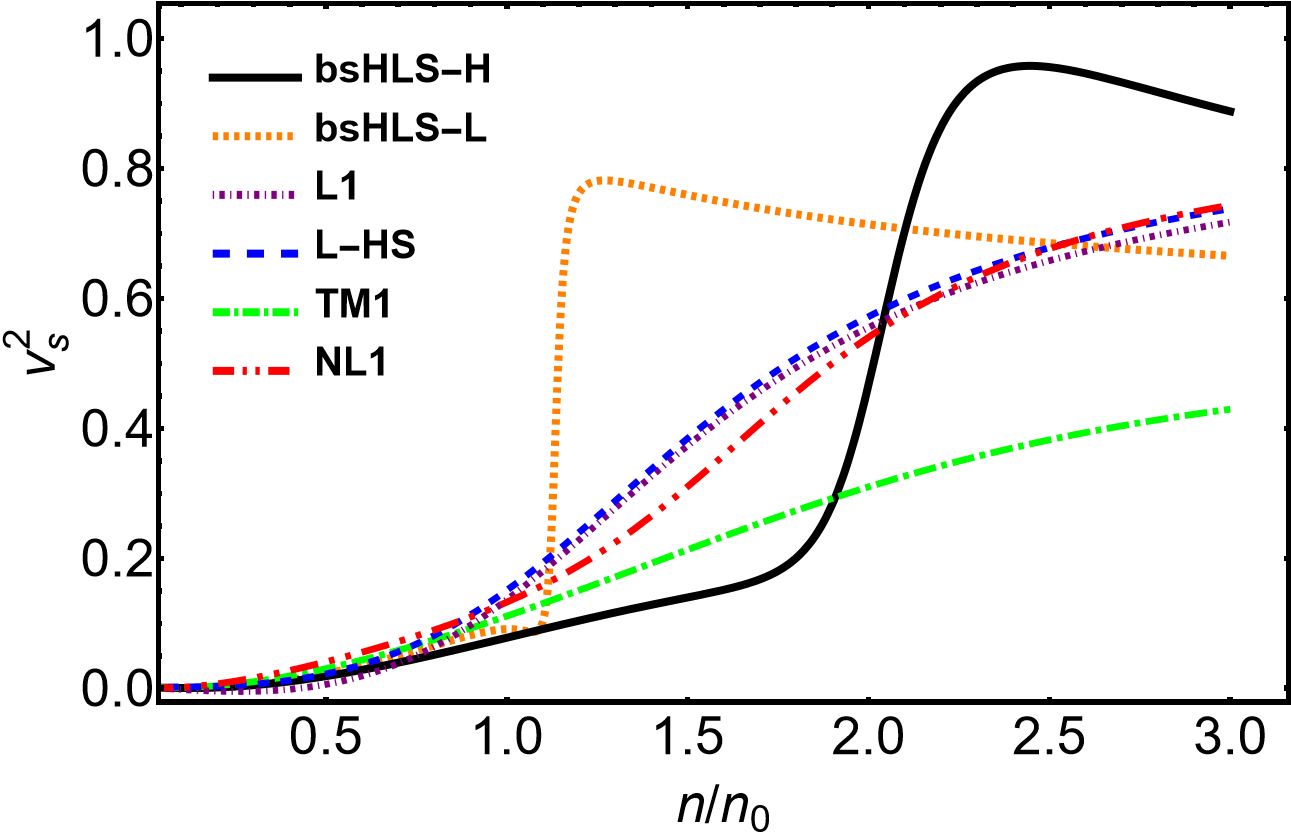}\quad \includegraphics[width=6.2cm]{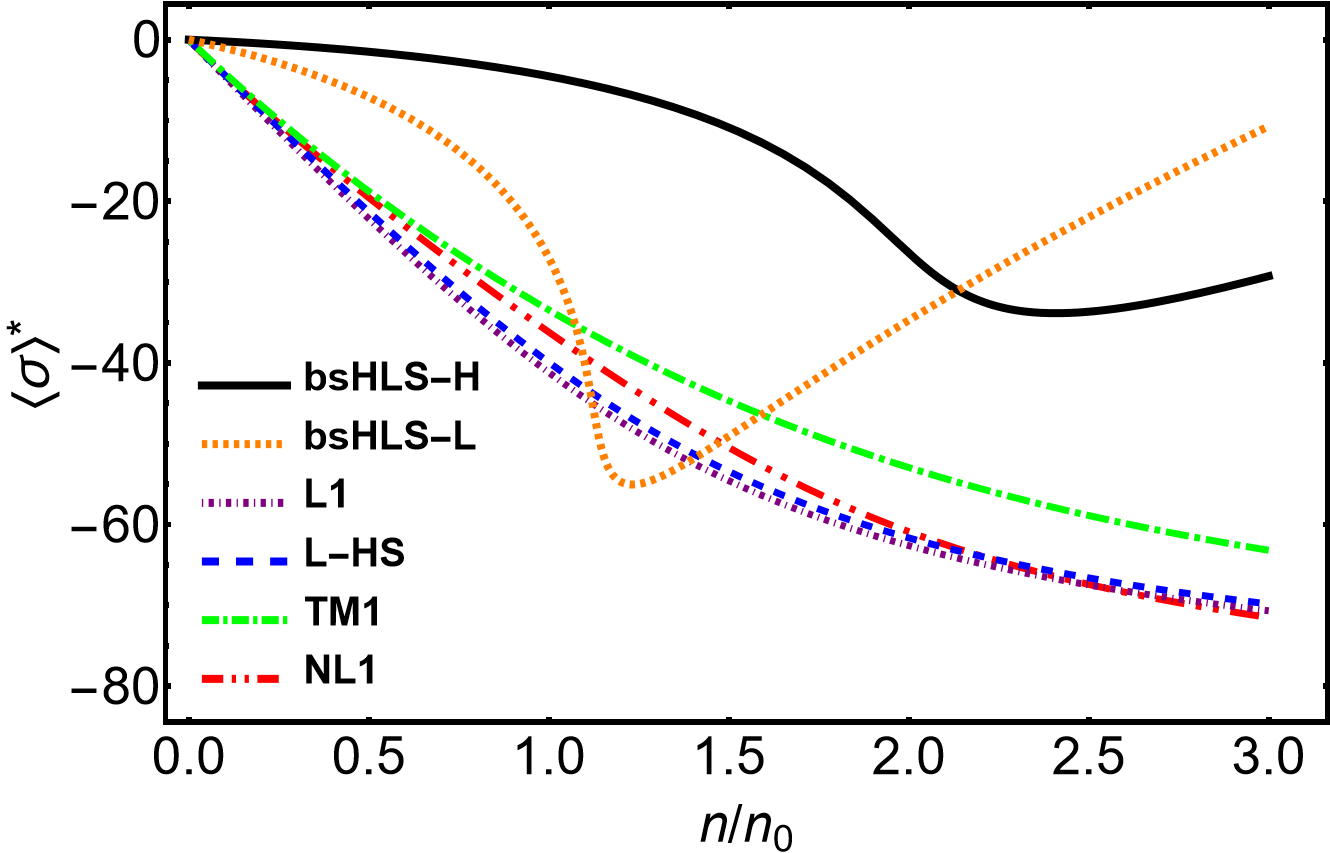}
	}
	\caption{Sound velocity (left panel) and VEV of sigma (right panel) vs density for from chiral scale EFT (bsHLS-H and bsHLS-L) and Walecka-type models. }
	\label{fig:peak}
\end{figure}

We next discuss the peak of SV which is normally attributed to the phase or configuration transition in the NM and has not been predicted in a unified hadronic model. Based on the chiral-scale EFT, by using the standard mean field approximation, we found that the peak of SV arises naturally~ \cite{Zhang:2024sju,Zhang:2024iye}. The results are shown in Fig.~\ref{fig:peak}.

Fig.~\ref{fig:peak} indicates that the origin of the peak is locked to the generic property of chiral-scale EFT that is not shared by the Walecka-type models. In the former, the effective screen mass of the vector mesons are modified by the dilaton compensator $\chi$, e.g., for the omega meson $ m_\omega^\ast = \frac{\langle \chi \rangle^\ast}{f_\chi} m_\omega $. The $m_\omega^\ast$, similarly for $\langle \chi \rangle^\ast$, cannot keep decreasing since otherwise the system cannot be stable due to the infinite omega meson repulsion. This behaviour is indeed demonstrated by the numerical calculation shown in Fig.~\ref{fig:peak}.

\section{Chiral-scale density counting rule}

We finally extend chiral-scale EFT to both finite density and temperature system. Since in the dilaton compensator, the meson field $\sigma$ is involved in $e^{\sigma/f_\chi}$, one should make a Taylor expansion. To estimate the contribution from each order of the expansion, we setup a counting rule validated to the density of the cores of massive stars---chiral-scale density counting (CSDC) rule---in addition to the chiral-scale counting rule in matter-free space. 

Since mesons can treated as the background fields induced by the nucleons in NM, their EOMs can be solved via the Green's function method with the corresponding fermion currents.
For example, for the $\sigma$ meson, 
\begin{equation}
	\label{eq:OGS}
	\sigma(x)=\int_V {\rm d}^4 x'D_{\sigma}(x-x')\left(-g_{\sigma NN}^{\rm OBE}\bar{N}(x')N(x')\right) ,
\end{equation}
where \(D_{\sigma}(x-x')\) is the retarded Green's function of the \(\sigma\) meson field. If the density does not go very high, the baryonic bilinear currents, like that in Eq.~\eqref{eq:OGS}, should have an order of \( n \propto k_{\rm F}^3\). For the magnitude of \(k_{\rm F}\) relevant to the cores of massive neutron stars, the characteristic momentum is up to \(k_{\rm{c}} \sim 700~{\rm MeV} \sim 10~n_0\), the same order as chiral-scale counting rules. 
Furthermore, the couplings are assumed to have a consistent order, e.g., the couplings between meson and nucleon is of  \(\mathcal{O}(p) \sim \mathcal{O}(k_c)\) except $g_{\pi NN}^{\rm{OBE}}$ due to its derivative coupling. 
In addition, the following assignment are made to the meson fields $\sigma,\ \pi,\ \omega_{\mu},\ \rho_{\mu}^i\ \sim \mathcal{O}(k_{\rm{c}}^2)$ for its derivative coupling with the baryons. For a detailed discussion, we refer to~\cite{Xiong:2025jxq}.

As an example, we show the NM properties calculated order by order in Tab.~\ref{tab:RES}. One can conclude that the CSDC rule works well and it is enough to consider the effect upto $\mathcal O(k_c^{12})$, the N$^4$LO. We argued in~\cite{Xiong:2025jxq} that the CSDC rule can also be applied to the thermal system up to temperature $100$~MeV. Here, we will not illustrate this extension.

\begin{table}[htb]
	\centering
	\caption{
		Nuclear matter properties around $n_0$. $E(n_0)$, $E_{\text {sym }}(n_0)$, $L(n_0)$, $K(n_0)$ are binding energy, symmetry energy, symmetry energy slope and incompressibility ~\cite{Danielewicz:2002pu} at $n_0$. 		\(T_{\rm c}\) is the GLPT critical temperature.
		$E(n_a)$ is binding energy at $n_a=1.5 n_0$. $n_0$ is in unit of fm$^{-3}$ and others are in unit of MeV.
	}
	\label{tab:RES}
	\begin{tabular}{lccccc}
		\hline
		\hline
		&Empirical & $\mathcal{O}(k_{\mathrm{c}}^{6})$ & $\mathcal{O}(k_{\mathrm{c}}^{8})$ & $\mathcal{O}(k_{\mathrm{c}}^{10})$ & $\mathcal{O}(k_{\mathrm{c}}^{12})$ \\
		\hline
		$n_0$ & $0.155 \pm 0.050$~\cite{Ma:2025llw} & $0.160$ & $0.160$ & $0.160$ & $0.158$ \\
		$E(n_0)$ & $-15.0 \pm 1.0 $~\cite{Sedrakian:2022ata} & $-16.3$ & $ -15.6$ & $-16.1$ & $-15.4$ \\
		$E(n_a)$ & $-13.3 \pm 0.5 $ ~\cite{Leifels:2015iei} & $-6.92$ & $-12.0$ & $-8.99$ & $-11.2$ \\
		$K(n_0)$ & $230 \pm 30$ ~\cite{Dutra:2012mb} & $566$ & $366$ & $572$ & $391$ \\
		$E_{\mathrm{sym}}(n_0)$ & $30.9 \pm 1.9 $~\cite{lattimer2013constraining} & $31.6$ & $32.1$ & $ 31.8$ & $32.2$ \\
		$L(n_0)$ & $52.5 \pm 17.5$~\cite{lattimer2013constraining} & $104$ & $83.2$ & $ 75.7$ & $77.8$ \\
		$T_{\rm c}$ & \(20.0\pm3.0\)~\cite{Karnaukhov:2003vp} & $19.0$ & $26.5$ & $24.0$ & $22.5$ \\
		\hline
		\hline
	\end{tabular}
\end{table}

\section{Summary and perspective}

In this contribution, we presented some predictions of the NM properties by using the chiral-scale EFT anchored on the chiral and scale symmetries of QCD, i.e., the pseudoconformal structure, the origin of SV and CSDC rules. 
In the future, we will make a full scan of the parameter space, higher loop corrections to NM and, extend the approach to include strange degree of freedom.


\bibliographystyle{unsrt}
\bibliography{ChiScaProRef}

\end{document}